\documentclass[aps,reprint,prl,floatfix]{revtex4-1}
\pdfoutput=1
\usepackage{amsmath,amssymb,subfigure,mathtools,relsize}
\usepackage{graphicx}
\usepackage{color}
\usepackage{floatpag}
\floatpagestyle{empty}
\usepackage{natbib}
\newcommand*\sq{\mathbin{\vcenter{\hbox{\rule{.9ex}{.9ex}}}}}

\begin{document}

\title{A Theory of Complex Stochastic Systems with Two Types of Counteracting Entities}

\author{Amin Zollanvari}
\affiliation{Nazarbayev University, Astana, Kazakhstan}


\begin{abstract}
Many complex systems share two characteristics: 1) they are stochastic in nature, and 2) they are characterized by a large number of factors. At the same time, various natural complex systems appear to have two types of intertwined constituents that exhibit counteracting effects on their equilibrium. In this study, we employ these few characteristics to lay the groundwork for analyzing such complex systems. The equilibrium point of these systems is generally studied either through the kinetic notion of equilibrium or its energetic notion, but not both. We postulate that these systems attempt to regulate the state vector of their constituents such that both the kinetic and the energetic notions of equilibrium are met. Based on this postulate, we prove: 1) the existence of a point such that the kinetic notion of equilibrium is met for the less abundant constituents and, at the same time, the state vector of more abundant entities is regulated to minimize the energetic notion of equilibrium; 2) the effect of unboundedly increasing less (more) abundant constituents stabilizes (destabilizes) the system; and 3) the (unrestricted) equilibrium of the system is the point at which the number of stabilizing and destabilizing entities increase unboundedly with the same rate. 
\end{abstract}

\maketitle
%
%


\section*{Introduction}

In 1970, Gardner and Ashby conducted a set of simulations experiments to study the stability of complex systems with entities that are connected at random \cite{gard}. In a seminal work, Robert May complemented their study by providing an analytical framework based on random matrix theory to describe the sharp transition from stability to instability as a function of the number of components (species) in the network \cite{May}. The analysis is based on a notion of stability known as neighborhood stability in a deterministic model. In particular, the model assumes that for a state vector $\mathbf{x}$ of size $n$ characterizing a small perturbation to the state (e.g., number of species) of each component around the equilibrium point, we have
\begin{equation}
\dot{\mathbf{x}}=(\mathbf{A}-\mathbf{I}_n)\mathbf{x}\,.
\label{may}
\end{equation}
In this model, $\mathbf{I}_n$ is an identity matrix of size $n$ that corresponds to the intrinsic stability of components, and $\mathbf{A}$ is the component-wise interaction matrix of size $n\times n$ with elements being random numbers drawn from a probability distribution with a mean of 0 and finite variance. One is then interested in the community equilibria, where all net growth rates of species are zero. Nevertheless, this analysis is limited in two ways: 

I) Although the interaction matrix $\mathbf{A}$ in equation\,({\ref{may}}) is random, the analysis does not capture the effect of  random environmental fluctuations. As May stated \cite{May}, ``Once the dice have been rolled to get a specific system, the subsequent analysis is purely deterministic". However, real systems are stochastic in nature, with underlying parameters exhibiting random fluctuations (see ref.\cite{MayBook} p. 17). 

II) Many complex systems seem to be the result of two types of constituents that have a counteracting influence on equilibrium of the system---for example, the the prey-predator ecological systems \cite{MayBook}, the neutron-proton model of a nucleus, or the stabilizing and destabilizing speculators in a financial system \cite{Frankel}. However, the states of all components of the model in equation\,({\ref{may}}) undergo the same machinery, which is to say that on the left of this equation, we have the net growth rate, and on the right we have the product of the state vector by a matrix. Therefore, it seems impossible that this model can capture the dynamics of such complex networks of interactions.

May later extended this framework to a fully stochastic ecological environment. In ref.\cite{MayBook}, he pictured an environment in which the interaction dynamics of species populations are not only random but are also subject to random environmental fluctuations. Assuming the random environmental fluctuations are characterized by a ``white noise", he formulated the dynamics of the problem as a multivariate Flokker-Planck equation. However, the complicated nature of these equations makes the exact solution of the multivariate setting hopeless. Instead, the exact solution in a univariate case (one species) can be determined, and some Gaussian approximations for the solutions of multivariate Flokker-Planck equations when the variance of white noise is very small are proposed. He concluded that the stability of the system (ref.\cite{MayBook}, p. 114) ``depends on the balance of power between the countervailing forces of stabilizing population interactions and randomizing environmental fluctuations''.  

In this work, we aim to establish a general theory that captures the dynamics of a complex stochastic system and quantifies the effect of two types of counteracting entities on the equilibrium point of the system. Our theory must account for complex systems in general and, as such, we may not rely on physical laws governing a specific field of study. Consequently, we establish the framework under some simple, yet general, conditions and postulates. The first natural question we need to answer is the following: What do we mean by a complex system?

To answer this question, we extend the definition offered by Freeman Dyson to characterize evolution from a complex nucleus to a complex system \cite{Dyson62I}; to wit, we refer to a system as a \textit{complex system} if it is characterized by many factors far too complicated to be understood in detail. Similar to Dyson's picture of a complex nucleus, we picture a complex system as a ``black box'' in which a large number of entities are
interacting according to unknown physical laws. 

In developing the framework, it is convenient to assume operators of finite but large dimensional space. In other words, rather than working with infinite dimensional operators in a Hilbert space, we approximate the complex system by discretization, keeping only part of the Hilbert space. This is the very first assumption Wigner made in developing the random matrix theory \cite{Wigner, Mehta}. In this discretized matrix picture, the action of an operator $\mathbf{H}$ in a vector space is characterized by a matrix product $\boldsymbol{\psi}=\mathbf{H}\boldsymbol{\phi}$ in which $\boldsymbol{\psi}$ and $\boldsymbol{\phi}$ are elements of the vector space. 

Throughout this article, we use boldface lowercase letters to denote column
vectors and boldface uppercase letters to denote matrices, with $\text{tr}[.]$ as
the trace operator.

\section*{Results}
\subsection*{The Dynamics of a Complex System}
\label{sec2}

We formalize the framework as follows: A complex system is composed of two types of counteracting entities, one working to stabilize the system and the other working to destabilize it. We refer to these stabilizing and destabilizing entities as SEs and DEs, respectively. These notions will be mathematically characterized later.

Suppose the set of SEs and DEs are constantly interacting. Let $\Omega$ be the state space. At any time, the state vector of these two types of entities is characterized by a nonzero $n$-dimensional random vector $\boldsymbol{\psi}(t,\omega)$ and a $p$-dimensional vector $\boldsymbol{\phi}(t,\omega)$ for $\omega \in \Omega$, but we do not know whether $\boldsymbol{\psi}(t,\omega)$ or $\boldsymbol{\phi}(t,\omega)$ represents the SEs or DEs' state vector. To capture the dynamics of the complex system, we propose the following stochastic system of equations that couples the state vector of SEs and DEs: 
\begin{equation}
\begin{aligned}
&\dot{\boldsymbol{\psi}}(t,\omega)=\Psi(t,\omega) [\mathbf{A}(t,\omega)\boldsymbol{\phi}(t,\omega)-\boldsymbol{\psi}(t,\omega)+\boldsymbol{\mu}(t,\omega)]\, \\
&\dot{\boldsymbol{\phi}}(t,\omega)=\Phi(t,\omega)\, [\mathbf{B}(t,\omega)\boldsymbol{\psi}(t,\omega)-\boldsymbol{\phi}(t,\omega)+\boldsymbol{\zeta}(t,\omega)]
\end{aligned}\,,
\label{system02}
\end{equation}
where $\mathbf{A}(t,\omega)$ and $\mathbf{B}(t,\omega)$ are $n\times p$ and $p \times n$ random matrices characterizing interactions between states $\boldsymbol{\psi}(t,\omega)$ and $\boldsymbol{\psi}(t,\omega)$, respectively, and matrices $\Psi(t,\omega)$ and $\Phi(t,\omega)$ are diagonal matrices with diagonal elements of vectors $\boldsymbol{\psi}(t,\omega)$ and $\boldsymbol{\phi}(t,\omega)$, respectively. Moreover, $\boldsymbol{\mu}(t,\omega)$ and $\boldsymbol{\zeta}(t,\omega)$ are considered to be multivariate white noise processes with each component having variance (in general, time dependent) $\gamma_{\boldsymbol{\mu},t}$ and $\gamma_{\boldsymbol{\zeta},t}$, respectively. For a full account of deriving the system of equations (\ref{system02}) from the Kolmogorov system of equations for population dynamics, see Supplementary Materials, Section I. Having the state vector $\boldsymbol{\phi}(t,\omega)$ in determining the state vector $\boldsymbol{\psi}(t,\omega)$ in (\ref{system02}), and vice versa, resembles the yin and yang nature of complex systems under study here. This philosophical view is not scientifically studied but as stated in ref.\cite{NEJ} ``...the whole is made up of the yin and
yang -- complementary, interdependent, and conceptually opposing entities that comprise a whole". Figure 1 symbolizes this philosophical belief in the context of a complex system with unboundedly many counteracting entities.

\begin{figure}[t!]
\centering
\includegraphics[scale=0.25]{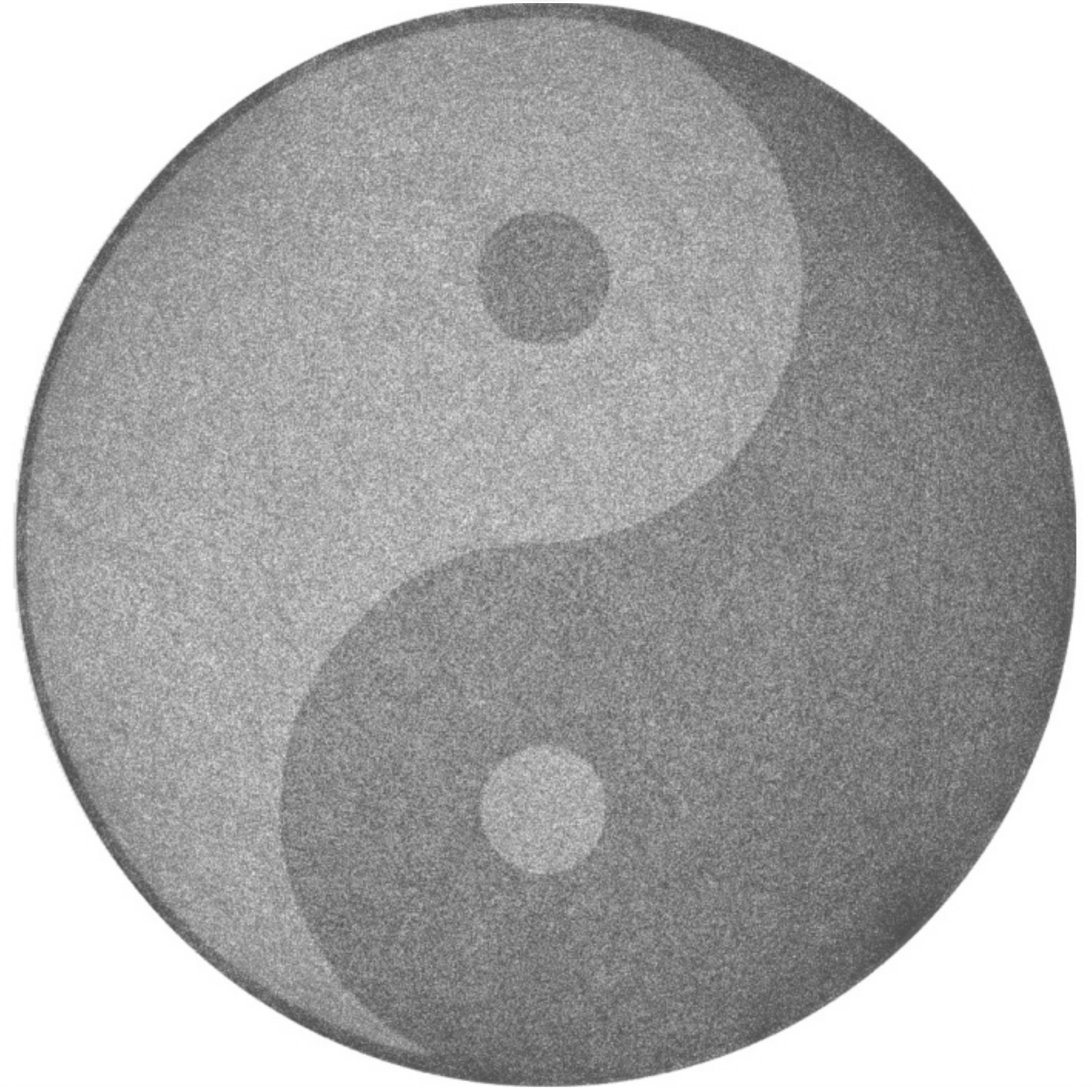}
\caption{The system of equations (\ref{system02}) resembles the symbolic yin and yang nature of a complex system with unboundedly many counteracting entities working towards an equilibrium point.}
\label{fig:SampleSignal}
\end{figure}

Modeling interactions by random matrices is similar to May's assumption on randomness of interactions in a complex ecological system \cite{May} and similar to assuming randomness of Hamiltonian in Schr\"odinger's equation \cite{Mehta}. The model proposed in equation ({\ref{system02}}) generalizes MacArthur's consumer-resource model with self-limitation to a fully stochastic environment (cf. \cite{MacArthur1970} Section V); in MacArthur's consumer-resource model, the vector of abundance of consumer and resource species and all other parameters are deterministic. For some special cases of equation ({\ref{system02}}) in stochastic settings (ref.\cite{MayBook}, Ch. 5 and p. 146).

\subsection*{Restricted Equilibrium}

Lotka describes two conceptions of equilibrium of a system--namely, a \textit{kinetic} and an \textit{energetic} conception of equilibrium (see ref.\cite{Lotka}, p. 143). 

\textit{\textbf{Kinetic notion of equilibrium:}} In this section, and for ease of notation, we omit $\omega$ from random vectors and matrices. From a kinetic perspective, equilibrium is a state at which certain velocities in an evolving system vanish. Using this notion, an equilibrium point is where $\dot{\boldsymbol{\psi}}(t)=\mathbf{0}_n$ and $ \dot{\boldsymbol{\phi}}(t)=\mathbf{0}_p$ (see ref.\cite{Lotka} p. 143 and ref.\cite{MayBook} p. 21). Having the zero velocity of state vector in equation ({\ref{system02}}) means
\begin{align}
&\dot{\boldsymbol{\psi}}(t)=\mathbf{0}_n \Leftrightarrow \boldsymbol{\psi}(t)=\mathbf{A}(t)\boldsymbol{\phi}(t)+\boldsymbol{\mu}(t)\,,
\label{system3}\\
&\dot{\boldsymbol{\phi}}(t)=\mathbf{0}_p\, \Leftrightarrow \boldsymbol{\phi}(t)=\mathbf{B}(t)\boldsymbol{\psi}(t)+\boldsymbol{\zeta}(t)\,.
\label{system31}
\end{align}
In this regard, we consider two cases:

\textbf{\textit{Case 1, $p<n$}}: Let us first consider equation ({\ref{system3}}). Given $\boldsymbol{\mu}(t)$ and $\mathbf{A}(t)$, we first fix $\boldsymbol{\phi}(t)$ on the right and try to solve this equation for $\boldsymbol{\psi}(t)$. Since we have $n$ unknown and $p$ equations, this is an underdetermined system of equations and generally has an infinite number of solutions. Now fix $\boldsymbol{\psi}(t)$ on the left and try to solve this equation for $\boldsymbol{\phi}(t)$. Since this is then an overdetermined system of equations, it is infinitely unlikely to have a solution (see ref.\cite{MacArthur1964}). To summarize, for $p<n$, this means that $\dot{\boldsymbol{\psi}}(t)=\mathbf{0}_n$ is not generally possible. Note that this argument does not hold for equation ({\ref{system31}}). Given $\boldsymbol{\zeta}(t)$ and $\mathbf{B}(t)$, if we fix $\boldsymbol{\psi}(t)$ on the right, $\boldsymbol{\phi}(t)$ is well determined. At the same time, if we fix $\boldsymbol{\phi}(t)$ on the left, we end up with an underdetermined system of equations with possibly an infinite number of solutions. To summarize, for $p<n$, the zero velocity principle only holds for equation ({\ref{system31}}). 
\begin{align}
&\dot{\boldsymbol{\psi}}(t)\neq\mathbf{0}_n, \label{PKKL}\\
&\dot{\boldsymbol{\phi}}(t)=\mathbf{0}_p.
\label{PKL}
\end{align}
This type of ``infinitely unlikely'' argument is inspired from MacArthur and Levins' arguments made in ref.\cite{MacArthur1964}.

\textbf{\textit{Case 2, $n<p$}}: Using a similar argument to the one in case 1, we see that the zero velocity conception of equilibrium only holds for equation ({\ref{system3}}), which leads to  
\begin{align}
&\dot{\boldsymbol{\psi}}(t)=\mathbf{0}_n,\\
&\dot{\boldsymbol{\phi}}(t)\neq\mathbf{0}_p.
\end{align}

\textit{\textbf{Energetic notion of equilibrium}}: The second conception is that a system is in equilibrium when certain functions having the dimensions of energy are \textit{minimum}. The question arises as to whether the kinetic and the energetic notions of equilibrium can coexist. In ref.\cite{MacArthur1970}, MacArthur showed that the state of an ecosystem at the kinetic equilibrium point of the competition equation is the point that minimizes a quadratic form, or, in other words, the energetic notion of equilibrium. Nevertheless, the competition equation considered in ref.\cite{MacArthur1970} is fully deterministic. Furthermore, an inherent assumption in the consumer-resource system of equations that he considered is that the number of resources and consumers is identical (and finite), while here we assume they increase unboundedly with an arbitrary asymptotic ratio. Therefore, we propose the following postulate:
\begin{quote}
\textbf{Postulate 1}: Complex systems attempt to simultaneously reach both the kinetic and the energetic equilibrium points.
\end{quote}

In this section, we assume the ratio of $p/n$ converges to a fixed asymptotic ratio $c$. In other words, the system is not allowed to change the limiting point of $p/n$. This assumption will be relaxed later to define an unrestricted point of equilibrium. Without loss of generality, let $p<n$, in which case the zero velocity of states only holds for $\boldsymbol{\phi}(t)$; thus, the system of equations ({\ref{PKKL}})-({\ref{PKL}}) holds. Since $\dot{\boldsymbol{\phi}}(t)=\mathbf{0}_p$ is possible, the system fixes values of ${\boldsymbol{\phi}}(t)$, or makes them independent of time (although still random w.r.t. $\omega$). At the same time, equation ({\ref{system31}}) dictates the relationship between ${\boldsymbol{\phi}}(t)$ and ${\boldsymbol{\psi}}(t)$. However, once values of ${\boldsymbol{\phi}}(t)$ are fixed, there are infinite solutions for ${\boldsymbol{\psi}}(t)$. From Postulate 1, the system moves the state vector ${\boldsymbol{\psi}}(t)$ (pick the solution) that corresponds to the minimum energy, that is, the inner product of difference between a functional of fixed ${\boldsymbol{\phi}}(t)$ and ${\boldsymbol{\psi}}(t)$. However, since the parameters of equation ({\ref{system02}}) are all random, the minimization of energy takes place on average over all realizations of random parameters. Mathematically, we can define this setting as
\begin{align}
\text{when}\;p<n:\notag\\[1ex]
\begin{split}
\underset{f_L}{\operatorname{min}}  \qquad & \mathlarger{{\Xi}}_{\psi,\phi,\mathbf{B},\boldsymbol{\zeta},n}(t)
\end{split}\label{opt}
\\[1ex]
\text{subject to}\qquad & \boldsymbol{\phi}(t)=\mathbf{B}(t)\boldsymbol{\psi}(t) + \boldsymbol{\zeta}({t})\;  \notag
\end{align}
where 
\begin{equation}
\begin{aligned}
&\mathlarger{{\Xi}}_{\psi,\phi,\mathbf{B},\boldsymbol{\zeta},n}(t) = \frac{1}{n}\big\langle \big(\boldsymbol{\psi}(t)-f_L(\boldsymbol{\phi}(t))\big), \big(\boldsymbol{\psi}(t)-f_L(\boldsymbol{\phi}(t))\big) \big\rangle= \\
&\frac{1}{n}  \text{E}_{\boldsymbol{\psi}(t),\boldsymbol{\phi}(t),\mathbf{B}(t),\boldsymbol{\zeta}({t})}\left[\big(\boldsymbol{\psi}(t)-f_L(\boldsymbol{\phi}(t))\big)^T\big(\boldsymbol{\psi}(t)-f_L(\boldsymbol{\phi}(t))\big)\right]\,
\end{aligned}
 \label{qazp}
\end{equation}
where $f_L(.)$ is assumed to be a linear mapping (\textbf{Postulate 2}). Since $\boldsymbol{\psi}(t)$ is an $n$-dimensional vector, a factor $\frac{1}{n}$ is used in equation ({\ref{qazp}}) to make the criterion an average inner product of differences per dimension. The linearity assumption used here is not an unreasonable assumption---it is at the core of various physical principles such the linearity of transformations from a rest frame to a moving frame in the special relativity. 

\subsection*{Restricted Equilibrium of A Complex System}
As described in the Introduction, rather than working with infinite dimensional spaces, for a complex system where both $p$ and $n$ are large, we study the limit of a finite dimensional problem. Based on equation ({\ref{qazp}}), determining the restricted equilibrium point of a complex system is equivalent to the following optimization problem:
\begin{align}
\text{when}\;p<n:\notag\\[1ex]
\begin{split}
\bar{\mathlarger{\Xi}}(t) \triangleq  \underset{f_L}{\operatorname{min}} \qquad &   \lim_{\substack{\\\textstyle{p\to\infty}\\\textstyle{p/n\to c}}}  \mathlarger{{\Xi}}_{\psi,\phi,\mathbf{B},\boldsymbol{\zeta},n}(t)
\end{split}
\\[1ex]
\text{subject to}\qquad & \boldsymbol{\phi}(t)=\mathbf{B}(t)\boldsymbol{\psi}(t) + \boldsymbol{\zeta}({t})\;  \notag
\label{optPP}
\end{align}

Using a similar argument, we can characterize the equilibrium of the system,
\begin{align}
\text{when}\;n< p:\notag\\[1ex]
\begin{split}
\bar{\mathlarger{\Xi}}(t) \triangleq \underset{f_L}{\operatorname{min}} \qquad &  \lim_{\substack{\\\textstyle{p\to\infty}\\\textstyle{p/n\to c}}}  \mathlarger{{\Xi}}_{\phi,\psi,\mathbf{A},\boldsymbol{\mu},p}(t)
\end{split}
\\[1ex]
\text{subject to}\qquad & \boldsymbol{\psi}(t)=\mathbf{A}(t)\boldsymbol{\phi}(t) + \boldsymbol{\mu}({t})\;  \notag
\label{nbm}
\end{align}
where $\mathlarger{{\Xi}}_{\phi,\psi,\mathbf{A},\boldsymbol{\mu},n}(t)$ is obtained from equation ({\ref{qazp}}) by exchanging $\phi$ and $\psi$ and replacing $\mathbf{B}$, $\boldsymbol{\zeta}$, and $p$ by $\mathbf{A}$, $\boldsymbol{\mu}$ and $n$, respectively. 

\begin{quote}
\textbf{Postulate 3}:  The elements of random matrices $\mathbf{A}(t)$, $\mathbf{B}(t)$ are i.i.d. random variables drawn from an arbitrary distribution with a finite mean and variance (in general, time dependent) of $\frac{1}{\sqrt{p}}$ and $\frac{1}{\sqrt{n}}$, respectively. Moreover, each of these elements is independent of every element of $\boldsymbol{\phi}(t)$ and $\boldsymbol{\psi}(t)$. 
\end{quote}

The choice of factor $\frac{1}{\sqrt{p}}$ is a common assumption in random matrix theory in order to make the variance of each row in the random matrix equal to 1 (see ref.\cite{couillet} p. 43). When $p\to\infty$ and $p/n\to c$, we show that (see Methods),
\begin{equation}
\begin{aligned}
&\bar{\mathlarger{\Xi}}(t)= \begin{cases} 
f(\gamma_{\boldsymbol{\zeta},t},c) - \frac{c-1}{2} \quad\quad\quad\,\;\; \text{if}\;\; c<1\\[2ex]
\frac{1}{c}\Big(f(\gamma_{\boldsymbol{\mu},t},c)+\frac{c-1}{2}\Big)\,\,
\quad\;\, \text{if}\;\; c>1
\end{cases}\,,
\label{jkl}
\end{aligned}
\end{equation}
where 
\begin{equation}
f(\gamma_{\boldsymbol{\zeta},t},c)= \frac{-\gamma_{\boldsymbol{\zeta},t}+\sqrt{(\gamma_{\boldsymbol{\zeta},t}-c+1)^2+4c\gamma_{\boldsymbol{\zeta},t}}}{2} \,.
\end{equation}

\textbf{Definition}: Any perturbation in the system that increases (decreases) the quantity $\bar{\mathlarger{\Xi}}(t)$ in  equation ({\ref{jkl}}) is a destabilizing (stabilizing) perturbation. In other words, any change in the system that increases (decreases) the total averaged asymptotic energy of the system is a destablizing (stablizing) effect. 

\textbf{Proposition 1}: Increasing the number of more abundant entities at a faster rate than the number of less abundant entities results in destabilizing the system. On the other hand, increasing the number of less abundant entities at a faster rate than the number of more abundant entities results in stabilizing the system.

\textit{Proof}: First, let $c<1$. After taking the derivative of $\bar{\mathlarger{\Xi}}(t)$ with respect to $c$ for a fixed $\gamma_{\boldsymbol{\zeta},t}$, it is easy to show that
\begin{equation}
\frac{\partial \bar{\mathlarger{\Xi}}(t)}{\partial c} = h(\gamma_{\boldsymbol{\zeta},t},c) -\frac{1}{2}<0\,,
\label{pppl}
\end{equation}
where 
\begin{equation}
h(\gamma_{\boldsymbol{\zeta},t},c)=\frac{-(\gamma_{\boldsymbol{\zeta},t}-c+1)+2\gamma_{\boldsymbol{\zeta},t}}{2\sqrt{(\gamma_{\boldsymbol{\zeta},t}-c+1)^2+4c\gamma_{\boldsymbol{\zeta},t}}}\,.
\end{equation}
Equation {\ref{pppl}} implies that increasing $c$ when $c<1$ (i.e., growing $p$, the number of less abundant entities, at a faster rate than $n$ when $\frac{p}{n} \rightarrow c$) has a stabilizing effect. When $c>1$, we can show that
\begin{equation}
\frac{\partial \bar{\mathlarger{\Xi}}(t)}{\partial c} >0\,,
\label{qwasd}
\end{equation}
The proof of equation ({\ref{qwasd}}) is not as straightforward as equation ({\ref{pppl}}) and is postponed to Supplementary Materials, Section II. Equation ({\ref{qwasd}}) implies that increasing $c$ when $c>1$ has a destabilizing effect. This also implies that increasing $n$ at a faster rate than $p$ has a stabilizing effect. $\sq$

Now we are in position to define the SEs and DEs in the system of equations ({\ref{system02}}). This is formalized in the following proposition. 

\textbf{Proposition 2}:
In the system of equations ({\ref{system02}}), when $p<n$, the state vector of SEs and DEs are represented by $\boldsymbol{\phi}(t,\omega)$ (the $p$-dimensional vector) and $\boldsymbol{\psi}(t,\omega)$ (the $n$-dimensional), respectively, and when $n<p$, the state vector of SEs and DEs are represented by $\boldsymbol{\psi}(t,\omega)$ and $\boldsymbol{\phi}(t,\omega)$, respectively. In other words, the state vector of SEs (DEs) is the one with a smaller (larger) dimension. 

\textbf{Proposition 3}: For a fixed $c<1$, increasing the variance of noise $\gamma_{\boldsymbol{\zeta},t}$ destabilizes the system. Similarly, when $c>1$, increasing the variance of noise $\gamma_{\boldsymbol{\mu},t}$ destabilizes the system

\textit{Proof}: The proof follows by fixing $c$ and taking the derivative of $\gamma_{\boldsymbol{\zeta},t}$ or $\gamma_{\boldsymbol{\mu},t}$ in equation ({\ref{jkl}}). $\sq$

\subsection*{Equilibrium of a Complex System}
The fundamental assumption in defining the restricted equilibrium point of the system used in the previous section is that the system has no flexibility to change the asymptotic relative abundance between SEs and DEs. Nevertheless, assuming a system with this additional flexibility, such as being able to change $c$, the energetic equilibrium point of the system is then the point that corresponds to the minimum energy w.r.t. $c$ as well. In fact it is hard not to believe that natural complex systems are capable of adjusting the relative abundance of their entities to evolve into an equilibrium state. 

In this regard, we define the \textit{unrestricted equilibrium} point, simply referred to as equilibrium, to be a point such that the objective function in equations ({\color{blue}{11}}) or ({\color{blue}{12}}) is minimized w.r.t. to both $f_L$ and $c$. With no prior knowledge of the inherent structure of the noise terms in the system of equation ({\ref{system02}}), we may assume $\gamma_{\boldsymbol{\mu},t}=\gamma_{\boldsymbol{\zeta},t}=\gamma_t$. That assumption leads to the following proposition.

\textbf{Proposition 4}: At each point in time, the system chacraterized 
by equation ({\ref{system02}}) reaches equilibrium when $c\rightarrow 1$. In other words, the system is at equilibrium when the number of SEs and DEs are asympotically equivalent. 

\textit{Proof}: The equilibrium point (unrestricted) of the system, if it exists, is characterized by the point where $\bar{\mathlarger{\Xi}}(t)$ is minimized with respect to $c$. For fixed $\gamma_t$, and from {Proposition 1}, we see that for $c<1$, $\bar{\mathlarger{\Xi}}(t)$ is a decreasing function of $c$, and the minimum occurs when $c \rightarrow 1^-$. That is,
\begin{equation}
\lim\limits_{c \rightarrow 1^-} \bar{\mathlarger{\Xi}} = g(\gamma_{t})\,,
\label{wq}
\end{equation}
where
\begin{equation}
g(\gamma_{t})=\frac{-\gamma_{t}+\sqrt{\gamma_{t}^2+4\gamma_{t}}}{2}\,.
\end{equation}
Similarly for $c>1$, $\bar{\mathlarger{\Xi}}(t)$ is an increasing function of $c$, and the minimum occurs when $c \rightarrow 1^+$,
\begin{equation}
\lim\limits_{c \rightarrow 1^+} \bar{\mathlarger{\Xi}} = g(\gamma_{t})\,.
\label{wq1}
\end{equation}
Equations ({\ref{wq}}) and ({\ref{wq1}}) yield,
\begin{equation}
\lim\limits_{c \rightarrow 1} \bar{\mathlarger{\Xi}} = g(\gamma_{t}) \,,
\label{wq1}
\end{equation}
which shows the existence of the equilibrium point when $c \rightarrow 1$. $\sq$

An interesting observation is that when $\gamma_{t}=c=1$, the ``energy'' of the system at equilibrium $\bar{\mathlarger{\Xi}}(t)$ approaches the golden mean, $ g(1) = 0.618033...\,$. 

Let us develop an intuitive understanding of the situation where $c \rightarrow 1$. Without loss of generality, consider the case where $p<n$, i.e., $c \rightarrow 1^-$. In this case, we have already seen that the kinetic notion of equilibrium only holds for $\dot{\boldsymbol{\phi}}(t)=\mathbf{0}_p$ (equations ({\ref{PKKL}}) and ({\ref{PKL}})); in other words, the state vector of SEs becomes time independent. At the same time, the state of DEs, ${\boldsymbol{\psi}}(t)$, is determined such that it minimizes the energetic notion of equilibrium. Nevertheless, when $c \rightarrow 1^-$, the system is approaching a point where the kinetic notion of equilibrium also holds for DEs (and at the same time, the energetic notion of equilibrium is minimized). In other words, the constraints on the state vector of DEs become tighter (the state vector of DEs becomes more and more time independent).

\section*{Discussion}
In this section, we discuss some of the implications of the proposed theory in economics and ecology.

\textbf{\textit{Stabilizing and Destabilizing Speculators:}} Here we consider the setting described by Frankel\cite{Frankel} (p. 178). We have a foreign exchange market with ``investors'' and ``spot traders''. When the value of the domestic currency exceeds its long-run equilibrium, investors will generally expect the value to depreciate, and as a result, they move to foreign currency, which drives the value of the domestic currency down. On the other hand, in a similar setting, spot traders buy more domestic currency because they expect its value continue to grow, which drives the value of the currency higher. There, Frankel presents a simplistic model of this system by defining some states as representing the fraction of world wealth allocated to domestic assets. He argues that instability in such an exchange market is either due to not having enough investors, or having too many spot traders\cite{Frankel}. 

Nevertheless, in such a setting where too many of either class create a destability of the market, it is natural to expect that when they are similarly abundant, the system approaches an equilibrium point. Let us formalize this system in terms of the system of equations ({\ref{system02}}). Suppose each element in vectors $\boldsymbol{\phi}(t,\omega)$ and $\boldsymbol{\psi}(t,\omega)$ represents the random fraction of world wealth allocated to domestic assets by each investor and each spot trader, respectively. Furthermore, let the association between $\boldsymbol{\phi}(t,\omega)$ and $\boldsymbol{\psi}(t,\omega)$ be characterized by matrices $\mathbf{A}(t,\omega)$ and $\mathbf{B}(t,\omega)$, with  i.i.d. elements being random variables with a finite mean and variance. In this case, we may couple the states of this system using equation ({\ref{system02}}) with $\gamma_{\boldsymbol{\mu},t}=\gamma_{\boldsymbol{\zeta},t}=\gamma_t$. Therefore, from Proposition 4, the system reaches an equilibrium point when $\frac{p}{n} \rightarrow 1$, or when the number of investors and spot traders are asymptotically equivalent. Nevertheless, contrary to Frankel's model, from Proposition 1 we conclude that whether investors are the SEs or DEs depends on their relative (asymptotic) abundance to spot traders. In other words, in a large complex system with many investors and spot traders, the more (less) abundant speculators are the DEs (SEs). 

\textbf{\textit{Prey-Predator Model:}} In ref\cite{May} (p.47-49), May presented a non-random $p$-predator-$n$-prey model of a complex ecosystem based on the Lotka-Volterra system of equations. Using an algebraic argument, he outlined the basis of ``the celebrated number of species equals number of resources theorem'' of MacArthur and Levins' \cite{MacArthur1964}. This ``theorem'' {per se}, has been a matter of long debate in the literature, and many authors perceive it as a tautology, not a principle \cite{Armstrong}. Nevertheless, as mentioned in the introduction, such a non-random system of equations does not capture the complexity of randomly fluctuating environment. Consider that each element of $\boldsymbol{\phi}(t,\omega)$ and $\boldsymbol{\psi}(t,\omega)$ denote the number of each predator and each prey, respectively. Similar to the speculator model, we assume random association matrices $\mathbf{A}(t,\omega)$ and $\mathbf{B}(t,\omega)$ and a coupling model {\ref{system02}} that characterize the relationship between prey and predators subject to some noise $\boldsymbol{\zeta}(t,\omega)$ and $\boldsymbol{\mu}(t,\omega)$. In this case, we see from Proposition 4 that the system has an equilibrium when $\frac{p}{n} \rightarrow 1$. Again, from Proposition 1, whether prey (or the predator) species are the SEs or DEs depends on the (asymptotic) relative number of various types of prey species ($n$) to predator species ($p$) in a complex stochastic ecosystem. 

\section*{Methods}
To prove equation (\ref{jkl}), we use: 1) Stieltjes transformation; 2) Mar\u{c}enko-Pastur law; and 3) orthogonality principle. For the readers' ease, the next section presents the Stieltjes transformation and the Mar\u{c}enko-Pastur law. This is then followed by the proof. 
\subsection*{Stieltjes Transformation and Mar\u{c}enko-Pastur law}
The Stieltjes transformation of a distribution function $F$ with density $f$ is defined as \cite{Bai:10},
\begin{equation}
\begin{aligned}
s_F(z)  = \int\limits_{-\infty}^{\infty} \frac{1}{\lambda-z} d F(\lambda), \quad z \in \mathbb{C}\backslash \text{Supp}(F) \,,
\end{aligned}
\label{pol}
\end{equation}
where $\text{Supp}(F) = \{x \in \mathbb{R} : f(x)>0\}$. Let $F_{\mathbf{G}_p}$ be the empirical spectral distribution of a $p \times p$ matrix ${\mathbf{G}_p}$ that is Hermitian, so that all the eigenvalues are real; to wit, 
\begin{equation}
\begin{aligned}
F_{\mathbf{G}_p}(x)  \triangleq \frac{1}{p} \sum_{i=1}^p \mathbf{1}_{\{\lambda_i(\mathbf{G}_p)\leq x\}}\,,
\end{aligned}
\label{pol1}
\end{equation}
where $\lambda_i(\mathbf{G}_p)$ are the eigenvalues of $\mathbf{G}_p$ and $\mathbf{1}_{\{.\}}$ is the indicator function. Applying $s_F(z)$ to $F_{\mathbf{G}_p}(x)$ yields,
\begin{equation}
\begin{aligned}
s_{F_{\mathbf{G}_p}}(z)  = \int\limits_{-\infty}^{\infty} \frac{1}{\lambda-z} d F_{\mathbf{G}_p}(\lambda) = \frac{1}{p}\text{tr}\left[\mathbf{G}_p-z\mathbf{I}_{p}\right]^{-1}\,.
\end{aligned}
\label{pol2y}
\end{equation}
Using the result of ref.\cite{Silver1995}, which is an extension of the so-called Mar\u{c}enko-Pastur law \cite{Marcenko}, there exists a unique distribution $\bar{F}$ with Stieltjes transformation $s_{\bar{F}}(z)$ such that
\begin{equation}
\begin{aligned}
\lim_{\substack{\\\textstyle{p\to\infty}\\\textstyle{p/n\to c}}} s_{F_{\mathbf{G}_p}}(z) \overset{a.s.}{\rightarrow } \lim_{\substack{\\\textstyle{p\to\infty}\\\textstyle{p/n\to c}}}  \text{E}_{\mathbf{G}_p}[s_{F_{\mathbf{G}_p}}(z)] \stackrel{}{\to}  s_{\bar{F}}(z)\,.
\end{aligned}
\label{polo}
\end{equation}
and
\begin{equation}
\begin{aligned}
s_{\bar{F}}(z) = \frac{1}{1-c-z-zcs_{\bar{F}}(z)}\,,
\end{aligned}
\label{poloik}
\end{equation}
that leads to the Stieltjes transformation of Mar\u{c}enko-Pastur density (ref.\cite{couillet} p. 51; also see ref.\cite{Bai:10} Theorem 3.7 for the case where the mean of each element in $\mathbf{G}_p$ is not necessarily zero). In the next section, we only need to work with the Stieltjes transformation of Mar\u{c}enko-Pastur density, but not with the density per se.

\subsection*{Derivation of Equation (\ref{jkl}) Using the Orthogonality Principle and the Random Matrix Theory}
We first consider the optimization problem presented in equation ({\color{blue}{11}}):
\begin{align}
\;p<n,\notag\\[1ex]
\begin{split}
\bar{\mathlarger{\Xi}}= \underset{f_L}{\operatorname{min}} \qquad &   \lim_{\substack{\\\textstyle{p\to\infty}\\\textstyle{p/n\to c}}}  \mathlarger{{\Xi}}_{\psi,\phi,\mathbf{B},\boldsymbol{\zeta},p}(t)
\end{split}
\\[1ex]
\text{subject to}\qquad & \boldsymbol{\phi}(t)=\mathbf{B}(t)\boldsymbol{\psi}(t) + \boldsymbol{\zeta}({t})\;  \notag
\label{optPP}
\end{align}
where $ \mathlarger{{\Xi}}_{\psi,\phi,\mathbf{B},\boldsymbol{\zeta},p}(t)$ is defined in (\ref{qazp}). Therefore, we can write,
\begin{equation}
\begin{aligned}
&\mathlarger{{\Xi}}_{\psi,\phi,\mathbf{B},\boldsymbol{\zeta},n}(t)=\\& \frac{1}{n}  \text{E}_{\boldsymbol{\psi}(t),\boldsymbol{\phi}(t),\mathbf{B}(t),\boldsymbol{\zeta}({t})}\left[\big(\boldsymbol{\psi}(t)-f_L(\boldsymbol{\phi}(t))\big)^T\big(\boldsymbol{\psi}_t-f_L(\boldsymbol{\phi}(t))\big)\right]\, \\
&=\,\frac{1}{n}\text{E}_{\mathbf{B}(t)}\Bigg[\\& \text{E}_{\boldsymbol{\psi}(t),\boldsymbol{\phi}(t),\boldsymbol{\zeta}({t})}\left[\big(\boldsymbol{\psi}(t)-f_L(\boldsymbol{\phi}(t))\big)^T\big(\boldsymbol{\psi}_t-f_L(\boldsymbol{\phi}(t))\big) \mid \mathbf{B}(t)\right] \Bigg]\\
&\stackrel{1}{=}  \text{E}_{\mathbf{B}(t)}\Big[\frac{\gamma_{\boldsymbol{\zeta},t}}{n} \text{tr}\left[\gamma_{\boldsymbol{\zeta},t}\mathbf{I}_{n} +   \mathbf{B}^{\dag}(t)\mathbf{B}^{}(t)\right]^{-1} \Big] ,
\end{aligned}
\label{qrds}
\end{equation}
where $\mathbf{B}^{\dag}(t)$ is the conjugate transpose of $\mathbf{B}^{}(t)$. Equality $\stackrel{1}{=}$ follows from (Bayesian) Gauss-Markov Theorem, which is a consequence of orthogonality principle (ref.\cite{kay} p. 391; also see equation (19) in ref.\cite{Silverstein}). Comparing the expression we have in bracket in (\ref{qrds}) with (\ref{pol2y}) and from (\ref{polo}) and (\ref{poloik}) we write
\begin{align}
&\bar{\mathlarger{\Xi}}=\lim_{\substack{\\\textstyle{p\to\infty}\\\textstyle{p/n\to c}}}   \text{E}_{\mathbf{B}(t)}\Big[\frac{\gamma_{\boldsymbol{\zeta},t}}{n} \text{tr}\left[\gamma_{\boldsymbol{\zeta},t}\mathbf{I}_{n} +   \mathbf{B}^{\dag}(t)\mathbf{B}^{}(t)\right]^{-1} \Big] \notag\\
& \stackrel{}{=} \lim_{\substack{\\\textstyle{p\to\infty}\\\textstyle{p/n\to c}}} {\gamma_{\boldsymbol{\zeta},t}}\times s_{F_{ \mathbf{B}^{\dag}\mathbf{B}}}(-{\gamma_{\boldsymbol{\zeta},t}}),
\label{r0}
\end{align}
where for simplicity of notations we omit dependency of $ \mathbf{B}(t)$ on $t$. At the same time, we have (see Lemma 3.1 in ref.\cite{couillet})
\begin{equation}
\begin{aligned}
& \frac{n}{p}s_{F_{\mathbf{B}^{\dag}\mathbf{B}}}(z)=s_{F_{\mathbf{B}\mathbf{B}^{\dag}}}(z)+\frac{p-n}{p}\frac{1}{z}.
\end{aligned}
\label{r00}
\end{equation}
On the other hand, 
\begin{equation}
\begin{aligned}
\lim_{\substack{\\\textstyle{p\to\infty}\\\textstyle{p/n\to c}}} s_{F_{\mathbf{B}\mathbf{B}^{\dag}}}(z) = \lim_{\substack{\\\textstyle{p\to\infty}\\\textstyle{p/n\to c}}} \frac{1}{p} \text{tr}\left[-z\mathbf{I}_{p} +  \mathbf{B}(t)\mathbf{B}^{\dag}(t)\right]^{-1}\\ \stackrel{1}{=}  \frac{1-c-z-\sqrt{(1-c-z)^2-4kz}}{2cz},
\end{aligned}
\label{r01}
\end{equation}
with $\stackrel{1}{=}$ following from (\ref{poloik}). Replacing (\ref{r01}) in (\ref{r00}), and then using the results in (\ref{r0}) we write,
\begin{equation}
\begin{aligned}
\bar{\mathlarger{\Xi}}= \frac{-\gamma_{\boldsymbol{\zeta},t}+\sqrt{(\gamma_{\boldsymbol{\zeta},t}-c+1)^2+4c\gamma_{\boldsymbol{\zeta},t}}}{2} - \frac{c-1}{2}.
\end{aligned}
\label{r02}
\end{equation}
The case of $n<p$ is simpler because we do not need to use (\ref{r00}). Using a similar machinery, we can show that in this case, 
\begin{equation}
\begin{aligned}
&\bar{\mathlarger{\Xi}} = \frac{-\gamma_{\boldsymbol{\mu},t}+\sqrt{(\gamma_{\boldsymbol{\mu},t}-c+1)^2+4c\gamma_{\boldsymbol{\mu},t}}}{2c} +\frac{c-1}{2c}.
\end{aligned}
\label{r1}
\end{equation}
Note that although the non-zero eigenvalues of $\mathbf{B}^{\dag}\mathbf{B}$ and $\mathbf{A}^{\dag}\mathbf{A}$ are the same, there is a difference between (\ref{r02}) and (\ref{r1}), which correspond to $p<n$ and $n<p$, respectively. This is because when $p<n$, $\mathbf{B}^{\dag}\mathbf{B}$ has $(n-p)$ additional 0 eigenvales each with a contribution of $\frac{1}{0-z}$ to the $s_{F_{\mathbf{B}^{\dag}\mathbf{B}}}(z)$. Additionally, when $p<n$, (\ref{qrds}) is multiplied by $\frac{1}{n}$ and when $n<p$, the counterpart equation is multiplied by a factor of $\frac{1}{p}$.



\end{document}